\documentclass[aps,showpacs,twocolumn,superscriptaddress,nofootinbib,floatfix,section 10pt]{revtex4-2}
\usepackage{amsfonts,amssymb,stmaryrd,latexsym,amsmath,braket}
\usepackage{graphicx,subfigure}
\usepackage{comment}
\usepackage{times}
\usepackage{slashed}
\usepackage{bm}
\usepackage{braket}
\usepackage{appendix}
\usepackage{enumitem}
\usepackage{multirow}
\usepackage{array}
\usepackage[colorlinks=true,backref=true,linktocpage=true,citecolor=blue,urlcolor=blue,linkcolor=blue,pdfpagemode=UseOutlines]{hyperref}
\usepackage[dvipsnames]{xcolor}

\begin{document}
\title{Practical limitations of the switching theorem for adiabatic state preparation
}
\author{Thomas D. Cohen}
\email{cohen@umd.edu}
\affiliation{Department of Physics and Maryland Center for Fundamental Physics, University of Maryland, College Park, MD 20742 USA}

\author{Andrew Li}
\email{enzuoli1002@gmail.com}
\affiliation{Montgomery Blair High School, Silver Spring, MD 20901 USA}

\author{Hyunwoo Oh}
\email{hyunwooh@umd.edu}
\affiliation{Department of Physics and Maryland Center for Fundamental Physics, University of Maryland, College Park, MD 20742 USA}

\author{Maneesha Sushama Pradeep}
\email{mpradeep@umd.edu}
\affiliation{Department of Physics and Maryland Center for Fundamental Physics, University of Maryland, College Park, MD 20742 USA}

\begin{abstract}

The viability of adiabatic quantum computation depends on the slow evolution of the Hamiltonian. The adiabatic switching theorem provides an asymptotic series for error estimates in $1/T$, based on the lowest non-zero derivative of the Hamiltonian and its eigenvalues at the endpoints. Modifications at the endpoints in practical implementations can modify this scaling behavior, suggesting opportunities for error reduction by altering endpoint behavior while keeping intermediate evolution largely unchanged. Such modifications can significantly reduce errors for long evolution times, but they may also require exceedingly long timescales to reach the hyperadiabatic regime, limiting their practicality. This paper explores the transition between the adiabatic and hyperadiabatic regimes in simple low-dimensional Hamiltonians, highlighting the impact of modifications of the endpoints on approaching the asymptotic behavior described by the switching theorem.

\end{abstract}

\date{\today}
\maketitle

\section{Introduction} \label{Sec:Introduction}

The adiabatic theorem, first introduced by Born and Fock in 1928~\cite{Born1928}, is one of the foundational results of quantum mechanics. It states that a quantum system remains in its instantaneous eigenstate if the Hamiltonian governing the system evolves sufficiently slowly and the initial state corresponds to an eigenstate. This theorem has far-reaching implications in diverse fields, ranging from atomic physics to quantum computing, where the gradual evolution of quantum states is a cornerstone of various state preparation protocols.

Adiabatic state preparation has emerged as a promising method for generating desired quantum states, particularly in the context of quantum computation and simulation. The method relies on initializing a system in a known ground state and evolving it adiabatically along a predefined path in the parameter space of the Hamiltonian, such that the system remains in the instantaneous ground state throughout the process. The efficacy of this approach hinges on the interplay between the evolution rate, the spectral properties of the Hamiltonian, and the constraints imposed by the adiabatic theorem.

Despite its utility, the practical application of the adiabatic theorem is often hindered by the need to balance efficiency and accuracy. Realistic implementations involve finite evolution times, leading to deviations from true adiabaticity and the accumulation of errors. Quantifying these errors and understanding their scaling behavior with evolution time, $T$, is crucial for determining the feasibility of adiabatic protocols in real-world scenarios.

In recent years, significant progress has been made in understanding corrections to the adiabatic theorem~\cite{PhysRevA.73.042104, PhysRevA.78.052508, PASSOS2020168172, Kolodrubetz:2017ofs, Sels:2017daz,  Bukov:2019fdp, PhysRevA.93.052107, Cohen:2024nbk, Cohen:2024kdi}. Tools such as the switching theorem~\cite{LENARD1959261, GARRIDO1962553, 10.1063/1.1704127, FJSancho_1966, Nenciu1981, Avron1987, JP, Nenciu1991, Nenciu1993, HAGEDORN2002235,  10.1063/1.3236685, PhysRevA.82.052305, 10.1063/1.4748968, PhysRevLett.116.080503, Cohen:2025riz} have provided insights into error propagation in the regime of asymptotically long evolution times.
The switching theorem can provide estimates for the errors using an asymptotic series in $1/T$ (where $T$ is the evolution time).
We define the hyperadiabatic regime, as described in~\cite{Cohen:2025riz}, as the range of $T$ where the asymptotic expansion is valid and primarily governed by its leading term.
As will be discussed in this paper, adiabatic behavior with small errors can occur outside of this hyperadibatic regime.

In the hyperadiabatic regime, the error is determined by the matrix elements of the lowest non-zero derivative of the Hamiltonian at its endpoints, scaled appropriately with a dimensionless time, while incorporating a relative phase factor that depends on the path.
In simple terms, the asymptotic scaling behavior for the error as a function of $T$, goes as $T^{-k}$ where $k$ is the lowest non-zero derivative of the time-dependent Hamiltonian $H(t)$ at either of its endpoints. Given this, it is natural to ask whether carefully engineering the instantaneous properties of the Hamiltonian at the initial and final points, while keeping the intermediate evolution largely unchanged, can significantly reduce the error. The switching theorem implies that in the hyperadiabatic regime the error will be reduced by powers of $1/T$. Viewed differently, the endpoint behavior of Hamiltonian paths in typical setups for adiabatic state preparation is essentially determined arbitrarily; the switching theorem suggests that if careful attention is given to selecting the endpoint behavior, errors could be significantly reduced. However, in order to take advantage of the switching theorem, the timescale must be long enough for the path to be in the hyperadiabatic regime. Thus, a key issue addressed in this paper is how large a timescale is needed for the hyperadiabatic regime.

This paper is organized as follows. Section~\ref{Sec:Switching} reviews the adiabatic theorem and the switching theorem. Section~\ref{Sec:Scaling} discusses different scaling behaviors as a function of $1/T$ for realistic systems and elucidates some subtleties associated with the practical implementation of an adiabatic evolution and the consequences of these for error estimation. In Section~\ref{Sec:Examples} and Section~\ref{Sec:Exp}, we compare the switching theorem estimates with some simple models that mock up some of the practical issues discussed in Section~\ref{Sec:Scaling}. Section~\ref{Sec:Discussion} discusses the implications of these results.

\section{Switching theorem} \label{Sec:Switching}


In this section, the adiabatic theorem and the switching theorem are briefly reviewed along with the relevant terminology.

In the real implementation of adiabatic state preparation, one cannot avoid errors from diabatic transitions because of the finiteness of the rate of evolution. If we define $U(t_f, t_i)$ as the time evolution operator of a given Hamiltonian between the initial and final time, $t_i$ and $t_f$, the error is written as
\begin{equation}
    \epsilon \equiv \left \lVert \left (1-|g_f\rangle \langle g_f |  \right) U(t_f, t_i)  |g_i \rangle \right \rVert, 
    \label{Eq:errordef}
\end{equation}
where $\lVert \cdot \rVert $ denotes the $L^2$-norm and $|g_i \rangle$, $|g_f \rangle$ are the ground states of the initial and final Hamiltonians, respectively.

The switching theorem describes the time evolution operator and the error in terms of an asymptotic series in $1/T$. Mathematically, the discrepancy between the adiabatically evolved state and the intended ground state of the Hamiltonian can be written as
\begin{subequations}
\begin{align}
    U |g_i \rangle \langle g_i |U^\dagger - |g_f \rangle \langle g_f | &\, = \,\sum_{n=1}^{n_{\max}(T)}\frac{B_n}{T^n} + R, \label{Eq:seriesOp} 
\end{align}
where $U \equiv U(t_f, t_i)$, and $B_n$ represents operators that are primarily determined by the properties of the trajectory's endpoints, with the intermediate path influencing the result only through phase factors that adjust the relative contributions from the endpoints. The remainder term, $R$, is an operator whose scaling with $T$ is slower than any power of $T$.  
$n_{\max}(T)$ indicates the highest number of terms in the asymptotic series that enhances the accuracy of the description for a given $T$.
From Eq.~(\ref{Eq:seriesOp}), the error has a similar asymptotic series in $1/T$:
\begin{align}
    \epsilon   &\, = \, \sum_{n=1}^{n_{\max}(T)} \frac{b_n}{T^n} + r,  \label{Eq:series}
\end{align}
\end{subequations}
$b_n$ and $r$ are coefficients that share similar properties to $B_n$ and $R$, respectively.

While the switching theorem provides an asymptotically accurate estimate the true error in Eq.~(\ref{Eq:errordef}), the error fluctuates a lot, which is the intrinsic property of the adiabatic theorem: rapid oscillations of phases cancel out excited states and only the ground state remains. In~\cite{Cohen:2025riz}, it was suggested to consider the timescale-averaged quantity for errors. The ``typical error'' is:
\begin{equation}
\bar\epsilon(T) \equiv  \frac{1}{2 \sqrt{T \tau_0}} \int_{T-\sqrt{T \tau_0}}^{T+\sqrt{T \tau_0}} d T' \,\epsilon(T')    \label{Eq:barerror}
\end{equation}
for $\tau_0>0$. In the large-$T$ limit, the typical error does not depend on $\tau_0$, and with this definition one can easily check the scaling behavior in $1/T$ without fluctuations.

If $H^m(0)=H^m(1)=0$ for all $m\leq n$, Ref.~\cite{Cohen:2025riz} showed that the typical error from the switching theorem in the hyperadiabatic regime gives
\begin{widetext}
\begin{equation}
    \bar\epsilon_n (T) = \frac{\bar b_n}{T^n} \; \; {\rm with} \; \;
    \bar{b}_n  =  \sqrt{ \sum_{j \neq g}   \left| \frac{ \langle j(0) | H^{(n)}(0) | g(0) \rangle}{\Delta_{j, g}^{n+1}(0)} \right|^2 + \sum_{j \neq g}\left| \frac{ \langle j(1) | H^{(n)}(1) | g(1) \rangle}{\Delta_{j, g}^{n+1}(1)} \right|^2  } 
    \equiv \sqrt{(\bar{b}_n^{0})^2 + (\bar{b}_n^{1})^2}  , \label{Eq:barbn}
\end{equation}
\end{widetext}
where the superscript $n$ on $H$ denotes the $n$th derivative of the Hamiltonian, and $|g(t)\rangle$ $|j(t)\rangle$ denotes the instantaneous ground state and the $j$th excited state, respectively. Additionally, $\Delta_{j, g}$ is the energy gap between the $j$th excited state and the ground state, i.e., $\Delta_{j, g}(t) = E_j(t) - E_g(t)$.


Moreover, in the hyperadiabatic regime, the typical error has the following inequality:
\begin{equation}
    \epsilon(T) \le \sqrt{2} \bar{\epsilon}(T),
\end{equation}
which was shown in~\cite{Cohen:2025riz}. Therefore, the hyperadiabatic regime is important in the sense that the error has a definite scaling behavior in $T$ and it can be bounded from the typical error.

Our goal in this paper is two-fold: first, we would like to determine the factors that determine the timescale $T$ at which the true error in an adiabatic state preparation setup is reasonably well approximated by the asymptotic scaling expressions from the switching theorem. Secondly, we would like to understand the circumstances in which manipulation of the endpoint properties of a Hamiltonian evolution (while leaving the intermediate evolution more or less unchanged) drastically reduces the error.

\section{Switching theorem in practice} \label{Sec:Scaling}
\subsection{Switching on of the switching theorem}

The irrelevance of the details of the intermediate evolution of the Hamiltonian in Eq.~(\ref{Eq:barbn}) provides a potential opportunity. If the system is in the hyperadiabatic regime, one could modify the instantaneous features of the Hamiltonian at its initial and final times while leaving the intermediate evolution largely the same and thereby reduce the errors. The key issue then becomes what controls whether the evolution is hyperadiabatic.

The switching theorem becomes particularly useful in the quantum computing context when the errors are dominated by the leading non-vanishing term in $1/T$ expansion in Eq.~(\ref{Eq:series}). For the scenario where $b_1\neq 0$, this is realized for $T$ values for which
\begin{equation}
b_1\, \geq b_2 \, T^{-1}, \, \cdots, \, b_{n}T^{-n+1}, \, \cdots.
\end{equation}
For a given set of $\{b_2\,,\,b_3\, ,\, \cdots\}$, this implies that the larger the value of $b_1$ (with the other $b_n$ fixed), the leading asymptotic behavior is realized for smaller values of $T$. This might naively suggest that increasing $b_1$ could be a way to observe the scaling behavior at smaller values of $T$. This is not sensible; increasing $b_1$ also increases the error, which is precisely the opposite of the goal.

Instead, the obvious way of reducing errors in the hyperadiabtic regime is by making the lower-order derivatives of the Hamiltonian zero at the endpoints while keeping the intermediate evolution largely unchanged. The switching theorem tells us that if $H^{(k)}(t_i)=H^{(k)}(t_f)=0$ for all $k\leq n$, then the asymptotic expression for the error scales as $1/T^{n+1}$. One would expect that increasing $n$ would reduce the error---provided that the system remains in the hyperadiabatic regime with that modification. However, there is a risk that such a modification could alter the timescale needed for the system to remain in the hyperadiabatic regime, making it impractically large. It is important to determine whether this happens.

To get information on this issue, this paper studies simple model Hamiltonians which are numerically tractable. What these studies show is that, as expected, at sufficiently long times, all of these models realize the asymptotic behavior dictated by the switching theorem. However, the timescale at which the processes become hyperadiabatic depends on some of the details of the intermediate evolution. In certain situations, this timescale can become very long. Moreover, situations may arise where the timescales are long enough to satisfy adiabatic conditions but not sufficient to enter the hyperadiabatic regime. In such cases, the errors can approximately follow a power-law scaling of $1/T^n$ where $n$ is {\it smaller} than the leading value predicted by the switching theorem. As the timescale increases and approaches the hyperadiabatic regime, the power-law behavior gradually converges to the value specified by the switching theorem.

\subsection{A subtlety}

To illustrate the subtle interplay between the timescale needed for hyperadiabatic behavior and the scaling behavior predicted by the switching theorem, it is useful to consider a simple example: 
start with $H_0(s)$, a generic Hamiltonian satisfying $H_0(0)=H_0(1)=0$, with a non-zero derivative at the endpoints: $H_0'(0),\, H_0'(1) \ne 0 $. Consequently, in the hyperadiabatic regime, $\bar{\epsilon}=\bar{b}_1/T$. Next, consider the function
\begin{equation}
  g(s;k)=\frac{s}{s+k}  
\end{equation}
where $k$ is a positive real number less than unity---and generally taken to be substantially less than unity. $g(s;k)$ has the properties that it is zero when $s=0$ and it approaches 1 when $s \gg k$. Thus, when $k\ll 1$, and $n$ is a positive integer $ H_{k,n}(s)$ defined by
\begin{equation}
    H_{k,n}(s) \equiv g(s;k)^{n} H_0(s) g(1-s;k)^{n} \label{Eq:H-modified}
\end{equation}
has the property that it is very close to $H_0(s)$ for most of the region $0<s<1$, but it alters the behavior at the endpoints such that the first $n$ derivatives of $H$ with respect to $s$ vanish. Moreover, the lowest non-vanishing derivative at the endpoints for $H_{k,n}(s)$ is given by 
\begin{equation}
H_{k,n}^{(n+1)}(0) = \frac{n! H_0'(0)} {k^n}, \; H_{k,n}^{(n+1)}(1) = \frac{n! H_0'(1)} {k^n}. \label{Eq:n-der}
\end{equation}

From Eqs.~(\ref{Eq:barbn}) and (\ref{Eq:n-der}), it is straightforward to see that in the hyperadiabatic regime, the typical error of the switching theorem is
\begin{equation}
    \bar{\epsilon}_n(T) = \frac{1}{T^n}\sqrt{\frac{n!}{k^n}} \sqrt{\sum_{j\neq g} \frac{\langle j(0) | H'(0) | g(0)\rangle}{\Delta^{n+1}_{j,g}(0)} + (0 \leftrightarrow 1)}. \label{Eq:nfac}
\end{equation}
Therefore, for a given $n$, in order for Eq.~(\ref{Eq:nfac}) to estimate the true typical error, the evolution should be in the hyperadiabatic regime, i.e., $\bar{\epsilon}_n(T) \ll 1$, which corresponds to $T \gg (n!)^{1/2n}/\sqrt{k}$. It demonstrates that as the modified Hamiltonian Eq.~(\ref{Eq:H-modified}) becomes close to the given Hamiltonian $H_0(s)$, i.e., $k\rightarrow 0$, it requires larger timescales to be in the hyperadiabatic regime. More importantly, as one modifies the Hamiltonian to have a better scaling behavior by increasing $n$, the timescale to be in the hyperadiabatic regime so that Eq.~(\ref{Eq:nfac}) can estimate the error precisely increases factorially.

Therefore, the core argument for applying the switching theorem is the following: even if one can manage the Hamiltonian path so that its first few derivatives at the endpoints are set to be zero while the modified Hamiltonian is nearly the same as the given Hamiltonian, one might not avoid large derivatives after such zero derivatives and the intended scaling behavior will appear only after large timescales.

\section{Examples} \label{Sec:Examples}

In this Section, we examine example setups in two-level and three-level systems where we can study the approach to the asymptotic scaling behavior described by the switching theorem. The examples we study have been constructed to mimic the subtleties of practical implementation that was discussed in Section~\ref{Sec:Scaling}. We start by defining a Hamiltonian $H_{0}(t)$, which we perturb at its endpoints to ensure that its first derivatives are zero at the endpoints while maintaining approximately the same intermediate evolution. We will then explore how modifications that deviate the actual Hamiltonian evolution from the intended evolution $H_0(t)$ influence the evolution of errors.

\subsection{Two-level system} \label{Sec:Two}

\begin{figure}[b]
    \centering
    \includegraphics[width=0.49\textwidth]{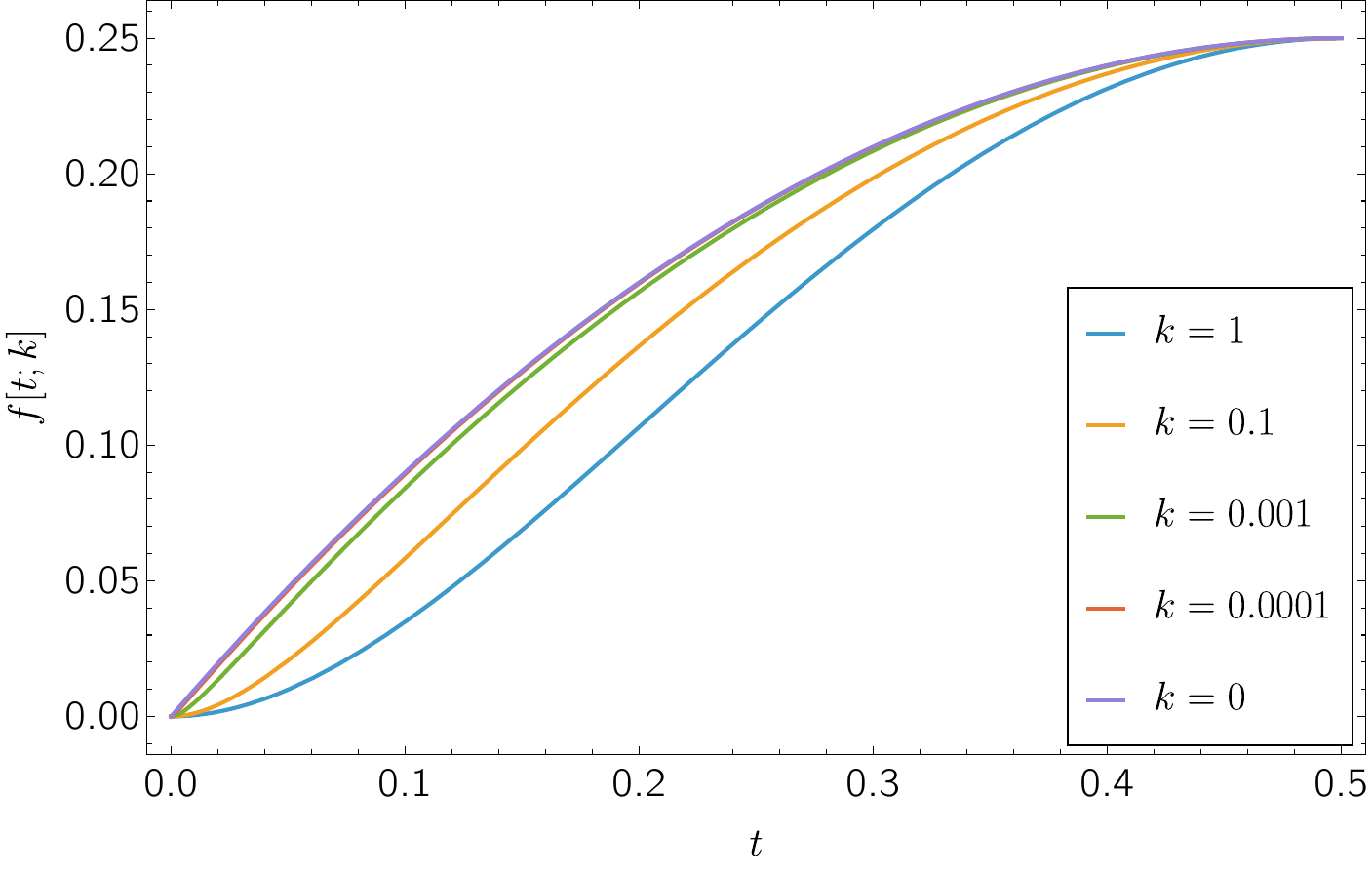}
    \caption{
    $f(t;k)$ with various $k$. $k=0$ corresponds to $f_0(t)$, whose first-order derivatives are not zero at the endpoints. }

    \label{fig:f}

\end{figure}

\begin{figure*}[t]
    \centering
    \includegraphics[width=0.98\textwidth]{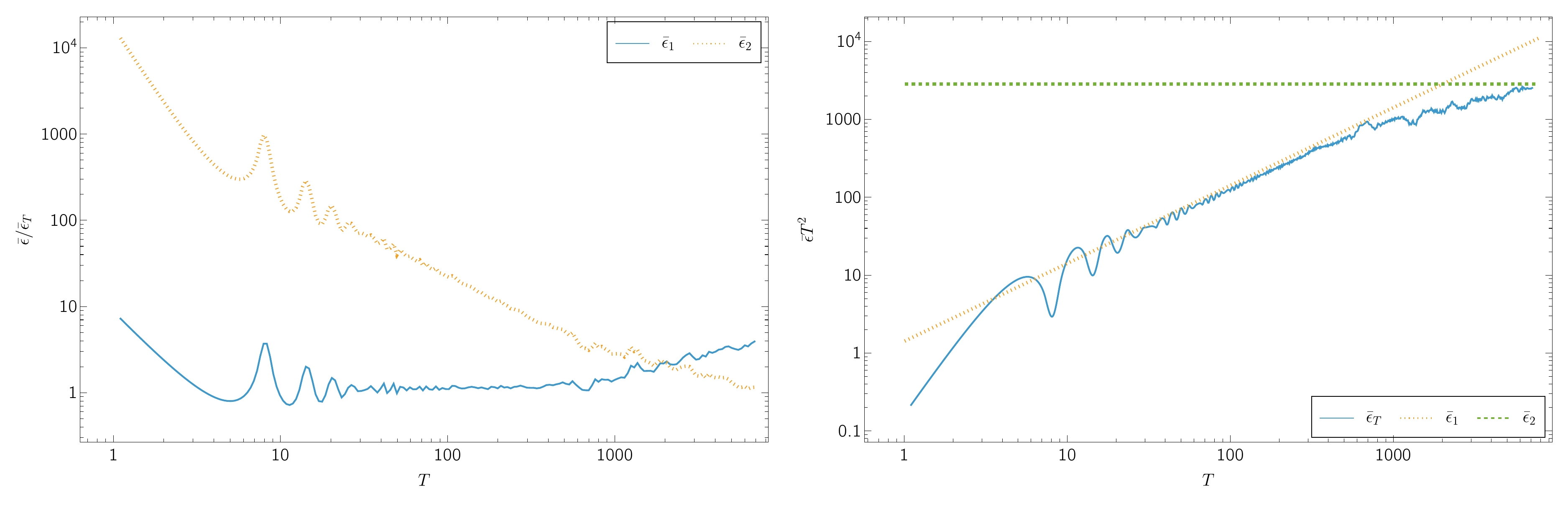}
    
    \caption{Comparison of errors, $\bar{\epsilon}_T$, $\bar{\epsilon}_1$, and $\bar{\epsilon}_2$, with $k=10^{-3}$ and $(E_0, E_1) = (1, 1)$. The left plot shows the ratio of switching errors and the averaged true error with the solid line being $\bar{\epsilon}_1$ and the dotted line being $\bar{\epsilon}_2$. The right panel displays the three average errors with respect to the timescale. Note that the y-axis for the right plot is $\bar{\epsilon}$ times $T^2$, not just the error $\bar{\epsilon}$, for visualizing the scaling behavior. 
    }

    \label{fig:2-level_smallk}

\end{figure*}

Consider a two-level system with the following Hamiltonian:
\begin{equation}
H_0(t) = 
    \begin{bmatrix}
    0 & E_1 f_0(t)  \\
    E_1 f_0(t)  & E_0 
    \end{bmatrix},  \label{Eq:H0_2level}
\end{equation} 
where $f_0(t)$ is given by:
\begin{equation}
    f_0(t) \equiv  t(1-t)  .
    \label{Eq:f0}
\end{equation}
Since $H^{'}_{0}(t)$ is not zero at the endpoints, the switching theorem implies that as $T\rightarrow\infty$, the typical error goes as $\bar{b}_1/T$ where $\bar{b}_1$ is given by Eq.~(\ref{Eq:barbn}). 

The model Hamiltonian that we study for adiabatic evolution may seem trivial in the sense that the initial and final states are the same. However, this feature is irrelevant for the questions that we want to answer, and the conclusions that follow would generally hold for non-trivial evolutions with different initial and final states, as is typical in realistic adiabatic state preparation setups. For presenting the plots, we will use $(E_0, E_1) = (1, 1)$ such that $\Delta_{1,0}=1$ in arbitrary units at the endpoints. In this work, we are not concerned about the explicit dependence of the energy gaps in the Hamiltonian\footnote{The focus of this work is on the relative delay in the onset of the asymptotic caused by endpoint modification, while the energy gaps are fixed at the endpoint.}. The true error is obtained by solving the Schr\"odinger equation\footnote{The time-dependent Schr\"odinger equation is solved using LSODA algorithm implemented on Mathematica.} and calculated using Eq.~(\ref{Eq:errordef}) and averaged using Eq.~(\ref{Eq:barerror}), which will be denoted by the symbol $\bar{\epsilon}_{T}$. We will then compare it with the switching theorem estimate $\bar{\epsilon}_n(T)$ in Eq.~(\ref{Eq:barbn}).

As elucidated by the series expansion for the error given by Eq.~(\ref{Eq:series}), if the first derivative of the Hamiltonian with respect to time vanishes at both endpoints, while the second derivative does not, then for $T\rightarrow\infty$, $\bar{\epsilon} \rightarrow \bar{b}_2/T^{2}$ where $\bar{b}_2$ is given by the expression in Eq.~(\ref{Eq:barbn}). 
Inspired by this observation, we ask whether we can modify the instantaneous properties of the Hamiltonian at its endpoint while keeping the intermediate behavior about the same, so as to reduce the error. For the Hamiltonian described in Eq.~(\ref{Eq:H0_2level}), we modify the time-dependent off-diagonal element such that its first-derivative vanishes by construction at the endpoints. As the reader would have anticipated, even local modifications of the function at the endpoints, would inevitably change the intermediate behavior. We consider modifications that do not change the intermediate behavior substantially.

Let us consider the following modified Hamiltonian: 
\begin{equation}
H(t; k) = 
\begin{bmatrix}
 0 & E_1 f(t; k)  \\
 E_1 f(t; k)  & E_0
\end{bmatrix} , \label{Eq:H_2level}
\end{equation}
where 
\begin{equation}
f(t; k) \equiv \left(\frac{1}{1+2k}\right)^2\frac{t}{t+k} t(1-t) \frac{1-t}{1-t+k}. 
\label{eq:fdefn}
\end{equation}
Here, $\frac{t}{t+k} \frac{1-t}{1-t+k}$ smoothens the behavior at the endpoints so that the first derivatives are zero when $k$ is not zero. The change in the intermediate path is small if $k$ is small. When $k$ is zero, the Hamiltonian is equal to $H_{0}(t)$, which, as discussed earlier, will show an error $\bar{\epsilon}\rightarrow \bar{b}_1/T$ as $T \rightarrow \infty$. The objective is to determine whether, for sufficiently large but reasonable values of $T$, this minor modification will lead to a reduction in error. The $t$-independent multiplicative factor ensures that $f(1/2,k)=f_0(1/2)$, i.e., the intermediate behavior is left as close as possible to the original Hamiltonian.

Fig.~\ref{fig:f} shows $f(t;k)$ for various $k$. Since it is even with respect to $t=0.5$, only its first half is plotted. It shows that when $k$ is small, $f(t;k)$ is nearly the same as $f_0(t)$ while its derivatives at the endpoints are 0. 
We want to stay as close as possible to the path traversed in the state space by $H_{0}(t)$ in the adiabatic limit, and therefore, $k$ is restricted to smaller values, which do not change the Hamiltonian path much.

Fig.~\ref{fig:2-level_smallk} compares the three errors, $\bar{\epsilon}_T$, $\bar{\epsilon}_1$, and $\bar{\epsilon}_2$, at fixed $k=10^{-3}$. Note that the error estimate $\bar{\epsilon}_1$ is made using the Hamiltonian $H_{0}(t)$ while the error estimate $\bar{\epsilon}_2$ is obtained using the Hamiltonian $H(t)$. The left plot shows the ratio of errors using the switching theorem, $\bar{\epsilon}_1$ and $\bar{\epsilon}_2$ to the averaged true error.
The right plot displays $\bar{\epsilon} T^2$ for averaged errors, providing a clearer visualization of the scaling behavior of errors.
Both panels show that when the timescale is small, $\bar{\epsilon}_1$ is a better approximation to the true error while in the asymptotic limit, the true error follows $\bar{\epsilon}_2$. Fig.~\ref{fig:2-level_smallk} nicely captures the approach of the true error to its asymptotic scaling form given by $\bar{\epsilon}_2$ for $k=10^{-3}$. As evident in the left plot, the ratio of $\bar{\epsilon}_1$ to the true error averages to 1 for smaller $T$ values, indicating that at smaller timescales $\bar{\epsilon}_1$ is a better approximation for the true error relative to $\bar{\epsilon}_2$. At larger timescales, one can see that the ratio $\bar{\epsilon}_1/\bar{\epsilon}_T$ deviates while $\bar{\epsilon}_2/\bar{\epsilon}_T$ begins to approach 1, thereby demonstrating the asymptotic scaling derived from the switching theorem. Similar conclusion can also be made from the right plot.

\begin{figure}[t]
    \centering
    \includegraphics[width=0.49\textwidth]{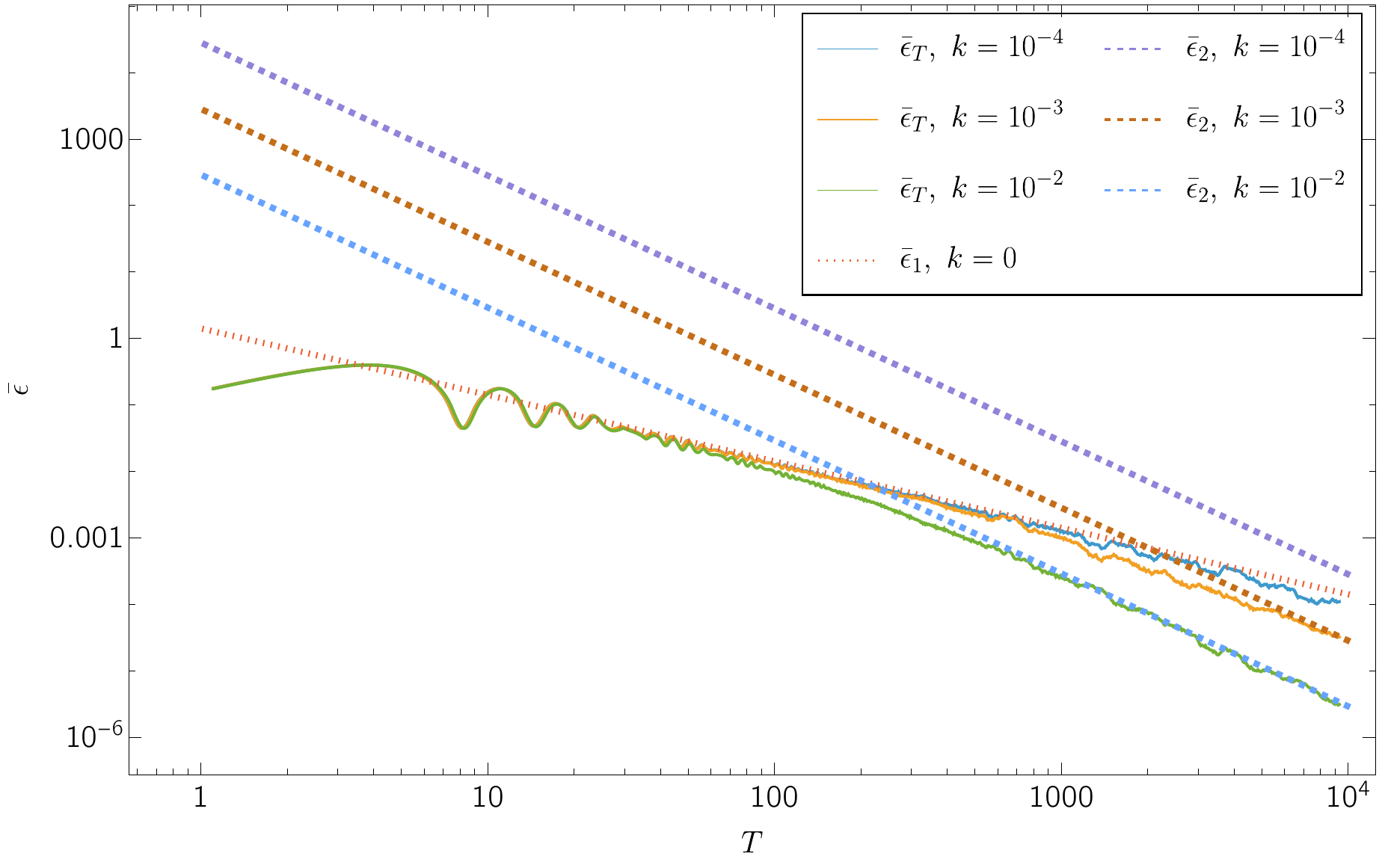}
    
    \caption{Comparison of average errors, $\bar{\epsilon}_T$, $\bar{\epsilon}_1$, and $\bar{\epsilon}_2$, with two $k$, $10^{-4}$ and $10^{-3}$, and $(E_0, E_1) = (1, 1)$. 
    Note that switching errors can be larger than 1 since they are estimation, but true errors, Eq.~(\ref{Eq:errordef}), are always less than 1.
    }
    \label{fig:2-level_variousk}

\end{figure}

Fig.~\ref{fig:2-level_variousk} shows errors for different values of $k$. Solid lines represent the averages of the true errors, the dotted line represents $\bar{\epsilon}_1$, and the dashed lines represent $\bar{\epsilon}_2$. For all cases, the average true errors follow $\bar{\epsilon}_1$ more closely at small timescales and $\bar{\epsilon}_2$ in the hyperadiabatic region, and $\bar{\epsilon}_1$ overestimates errors eventually. However, the scale of hyperasymptotic limits for true errors are different based on the size of $k$. We note the following: for each $k$ value shown in the plot, the error at lower timescales is better described by the switching theorem estimate obtained using the Hamiltonian $H_{0}(t)$ and therefore the true error shows $1/T$ behavior for smaller values of $T$. For larger timescales, i.e,  $T\gg O(k^{-1}\Delta^{-1}_{1,0})$, the true errors approach the switching theorem scaling estimate ($\bar{\epsilon}_T \approx \bar{\epsilon}_2 \approx T^{-2}$) obtained using the modified Hamiltonian, $H(t)$.
For smaller $k$, the true errors follow $\bar{\epsilon}_1$ longer and finally follow $\bar{\epsilon}_2$ in large $T$, which corresponds to small true errors compared to cases with larger $k$. 
In general, we find that as $k\rightarrow 0$---meaning the intermediate evolution with $H(t)$ becomes closer to that with $H_0(t)$ while the endpoint behavior becomes more sudden---the timescale required for the true error to approach $\bar{\epsilon}_2$ for $H(t)$ also increases. 
This analysis demonstrates that for sufficiently large values of $T$, the asymptotic scaling of the error in powers of $1/T$ is indeed achieved. However, the hyperadiabatic timescale depends upon the relative magnitude of the lower-order derivatives at the endpoints. It indicates that in order to make remarkable improvements such as by order of magnitude with practical values of $T$, minor modifications at the endpoints might not suffice.

\subsection{Three-level system}

We analyzed how a two-level system approaches the asymptotic behavior predicted by the switching theorem. We expect to see similar behavior in systems with more than one accessible excited states. In this Section, we do a similar analysis with a three-level system. The Hamiltonian $H_{0}(t)$, analogous to the two-level case, is defined as follows:
\begin{equation}
H_0(t) = 
\begin{bmatrix}
 1 & E_1 f_0(t) & E_2 f_0(t) \\
 E_1 f_0(t) & 2 &  E_3 f_0(t) \\
 E_2 f_0(t) & E_3 f_0(t) & 3
\end{bmatrix},  \label{Eq:H0_3level}
\end{equation}
where $f_0(t)$ is defined in Eq.~(\ref{Eq:f0}).
We can make modifications similar to the previous case:
\begin{equation}
H(t; \mathbf{k}) = 
    \begin{bmatrix}
    1 & E_1 f(t; k_1) & E_2 f(t; k_2) \\
    E_1 f(t; k_1) & 2 &  E_3 f(t; k_3) \\
    E_2 f(t; k_2) & E_3 f(t; k_3) & 3
    \end{bmatrix} , \label{Eq:H_3level}
\end{equation}
where $f(t; k)$ is defined in Eq.~(\ref{eq:fdefn}). As is evident, $H(t, \mathbf{k})\rightarrow H_{0}(t)$ as $\mathbf{k}\equiv (k_1,k_2,k_3) \rightarrow \mathbf{0}$ and $\frac{\partial H}{\partial t}(t_i, \mathbf{k})=\frac{\partial H}{\partial t}(t_f, \mathbf{k})=0$ when $\mathbf{k}\neq 0$. 

We consider two cases for $\mathbf{k}$ and $\mathbf{E}=(E_1,E_2,E_3)$:
\begin{itemize}
    \item Case 1: $E_1=E_2=E_3=1$ and $\mathbf{k}=(k,k,k)$:  
    \begin{equation}
    H(t; \mathbf{k}) = 
    \begin{bmatrix}
     1 &  f(t; k) &  f(t; k) \\
      f(t; k) & 2 &   f(t; k) \\
      f(t; k) & f(t; k) & 3
    \end{bmatrix} . \label{Eq:H_3levelex1}
    \end{equation}

    \item Case 2 : $(E_1, E_2, E_3) = (1, 0, 1)$ and $\mathbf{k} = (k, 0, 0)$: 
    \begin{equation}
    H(t; \mathbf{k}) = 
    \begin{bmatrix}
     1 & E_1 f(t; k_1) & 0 \\
     E_1 f(t; k_1) & 2 &  E_3 f_0(t) \\
     0 & E_3 f_0(t) & 3
    \end{bmatrix}. \label{Eq:H_3levelex2}
\end{equation}
\end{itemize}

Notice that Case 1 may be considered as a straightforward generalization of the two-level system that we discussed before. Diabatic transitions between any levels are suppressed to the first-order in $1/T$ for this Hamiltonian. Case 2, although similar to the previously discussed two-level system, has distinctive characteristics. As one may notice, adiabatic transitions between the ground state to any of the excited states up to first-order in $1/T$ are suppressed for $k\neq 0$. However, adiabatic transitions between the excited states are allowed at the first-order in $1/T$. Note that since the switching error formulas, Eq.~(\ref{Eq:barbn}), only consider transitions between the ground state and other states and there is no transition between the ground state and the second excited state at the first-order in $1/T$, the first non-trivial switching error is $\bar{\epsilon}_2$ for the ground state when $k \neq 0$.

We begin our analysis with Case 1, exploring different values of $k$ and comparing the true error with the switching error estimates, $\bar{\epsilon}_1$ and $\bar{\epsilon}_2$. $\bar{\epsilon}_1$ and $\bar{\epsilon}_2$ are calculated as before with Hamiltonians $H_0(t)$ and $H(t)$ respectively.

\begin{figure}[t]
    \centering
    \includegraphics[width=0.49\textwidth]{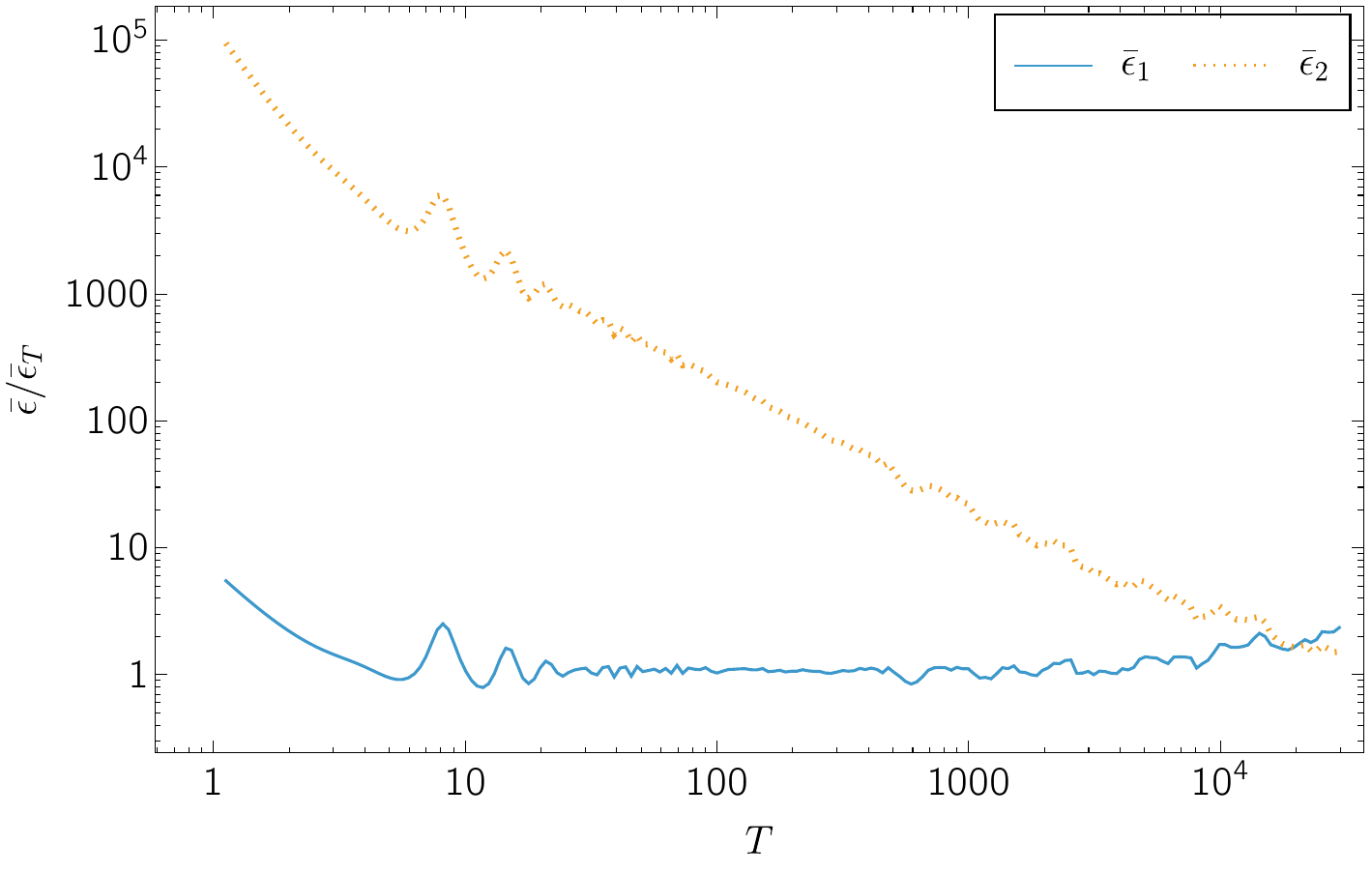}
    
    \caption{Comparison of average errors, $\bar{\epsilon}_1$, and $\bar{\epsilon}_2$, with respect to $\bar{\epsilon}_T$,  for the three-level system with identical off-diagonal elements at a fixed $k_1 =k_2 =k_3 = 10^{-4}$, while varying $T$.
    }

    \label{fig:3-level_smallk}

\end{figure}

\begin{figure}[t]
    \centering
    \includegraphics[width=0.49\textwidth]{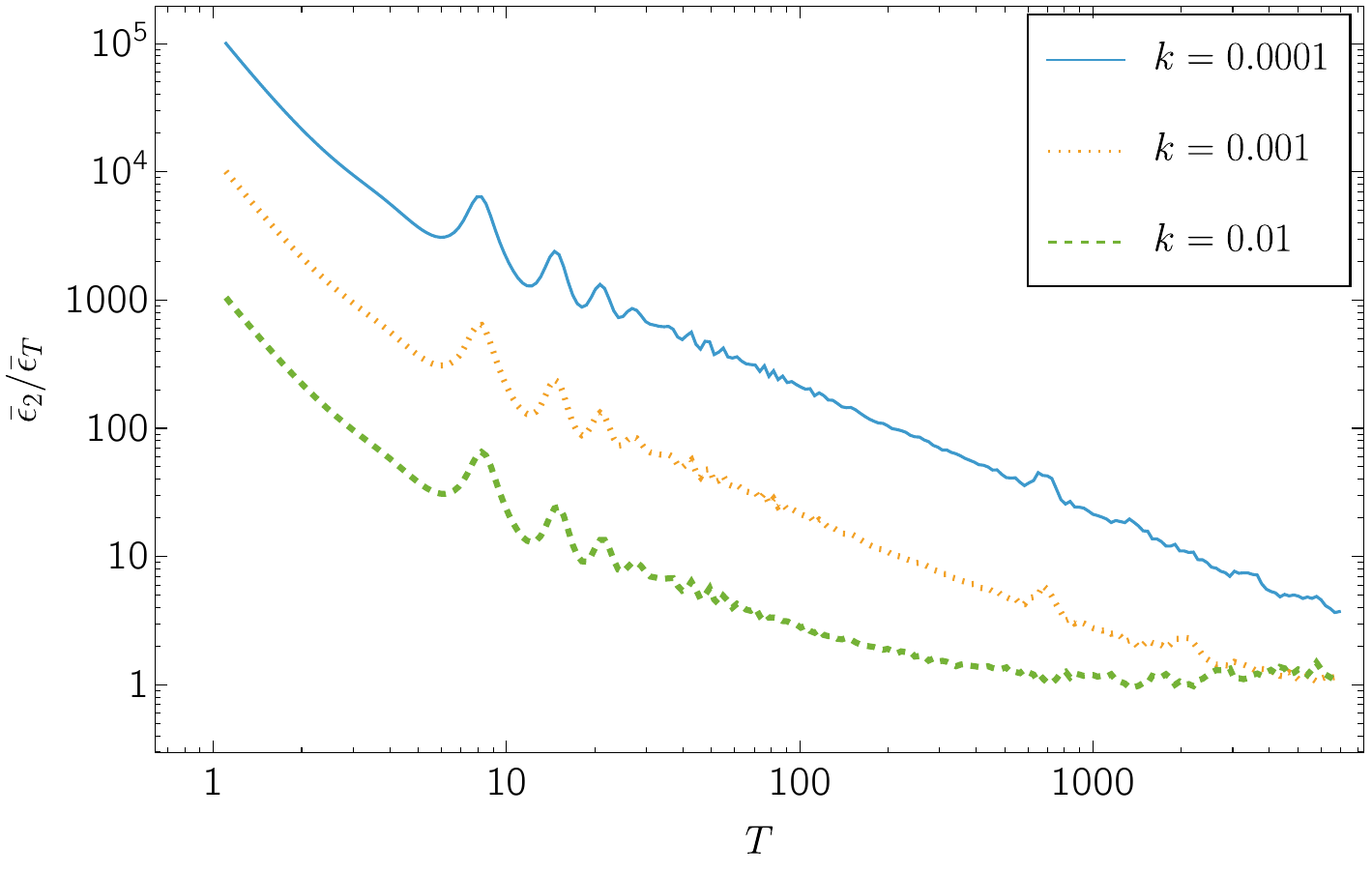}
    
    \caption{Comparison of the error ratio, $\bar{\epsilon}_2$/$\bar{\epsilon}_T$, by changing $k$ for the three-level system with identical off-diagonal elements, Eq.~(\ref{Eq:H_3levelex1}).
    }

    \label{fig:3-level_variousk}
\end{figure}

Fig.~\ref{fig:3-level_smallk} compares $\bar{\epsilon}_1$ and $\bar{\epsilon}_2$ with respect to the true errors at $k=10^{-4}$. As the two-level system, $\bar{\epsilon}_2$ is far from $\bar{\epsilon}_T$ with small $T$ but $\bar{\epsilon}_T$ converges to $\bar{\epsilon}_2$ in the large-$T$ limit with similar reasons. 
As before, $\bar{\epsilon}_1$ is calculated with the original Hamiltonian defined in Eq.~(\ref{Eq:H0_3level}).

In Fig.~\ref{fig:3-level_variousk}, the average error ratios $\bar{\epsilon}_2$/$\bar{\epsilon}_T$ are compared for different $k$. It shows that for large $k$, the average true errors follow $\bar{\epsilon}_2$ even at small $T$, whereas for small $k$, they gradually approach $\bar{\epsilon}_2$ as the timescale increases, similar to Fig.~\ref{fig:2-level_variousk}.

\begin{figure}[t]
    \centering
    \includegraphics[width=0.49\textwidth]{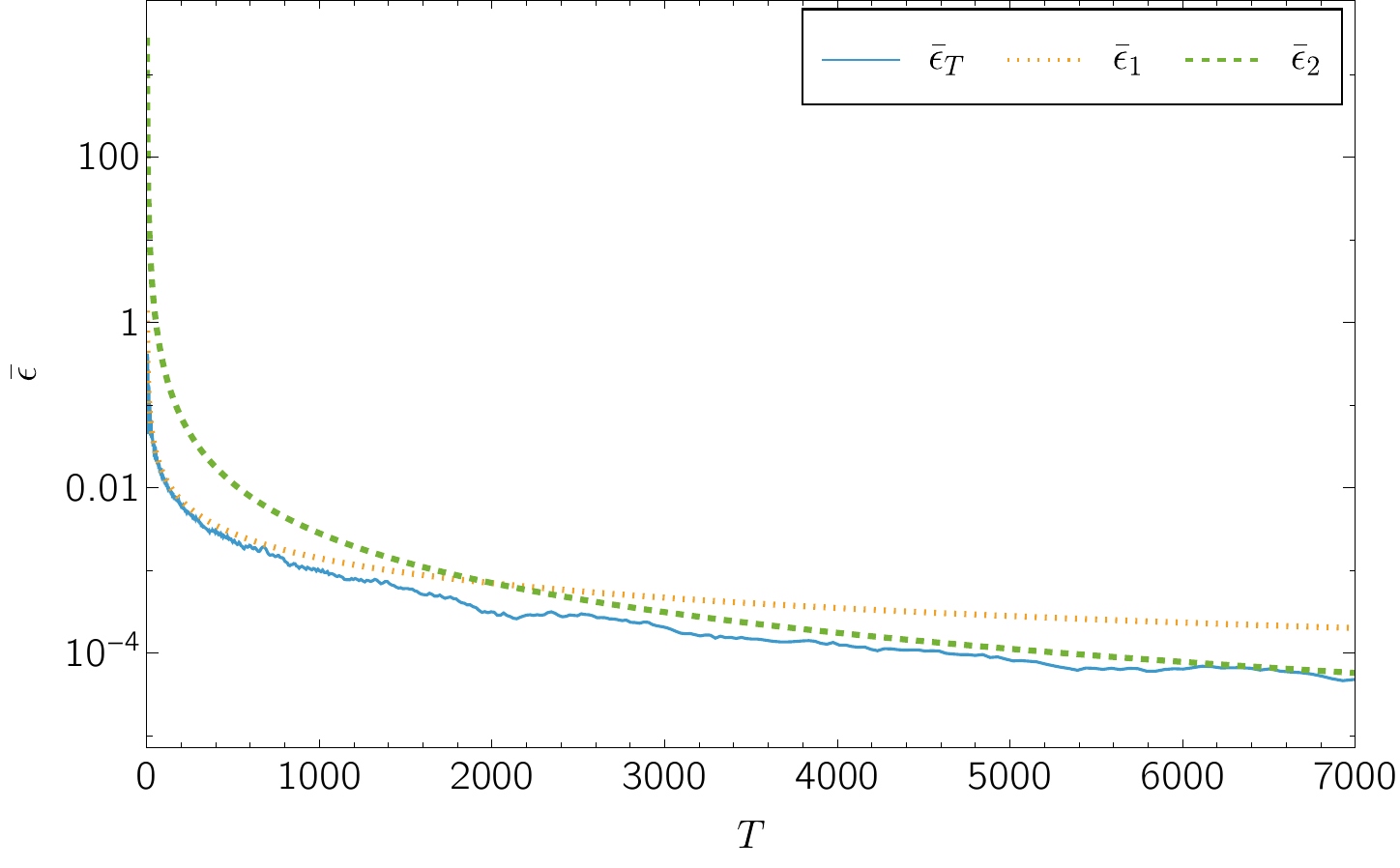}
    
    \caption{Three average errors, $\bar{\epsilon}_T$, $\bar{\epsilon}_1$, and $\bar{\epsilon}_2$, are shown. Parameters in Eq.~(\ref{Eq:H_3level}) are chosen as $(E_1, E_2, E_3) = (1, 0, 1)$ and $\mathbf{k} = (10^{-3}, 0, 0)$. Note that for the Hamiltonian with given parameters, $\bar{\epsilon}_1$ is zero, so the $\bar{\epsilon}_1$ in the plot is calculated using Eq.~(\ref{Eq:H0_3level}).
    }

    \label{fig:3-level_diff_fixedk}
\end{figure}

\begin{figure}[t]
    \centering
    \includegraphics[width=0.49\textwidth]{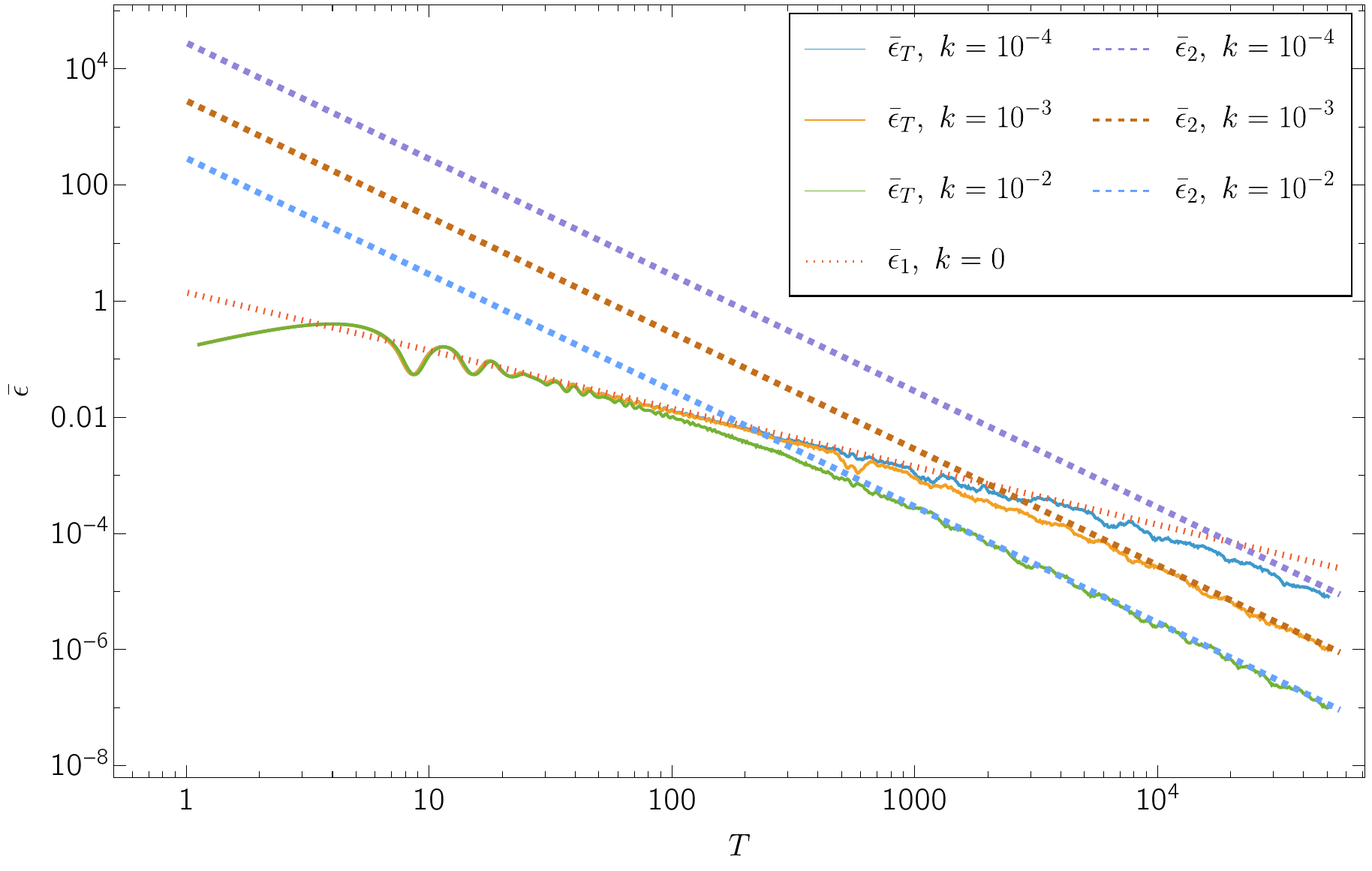}
    
    \caption{Average errors for three-level systems with different off-diagonal elements are displayed. In Eq.~(\ref{Eq:H_3level}), we choose $(E_1, E_2, E_3) = (1, 0, 1)$ and $ \mathbf{k} = (k, 0, 0)$ and $k$ varies. 
    }

    \label{fig:3-level_diff_diffk}
\end{figure}

We now analyze Case 2 using the same tools applied to previous systems. Fig.~\ref{fig:3-level_diff_fixedk} illustrates that for small $T$, the true errors follow $\bar{\epsilon}_1$, while for large $T$, it follows the expected $\bar{\epsilon}_2$, consistent with previous systems.

As in previous examples, $\bar{\epsilon}_T$ converges to $\bar{\epsilon}_2$ as $T$ increases, while for small $T$, $\bar{\epsilon}_T$ follows $\bar{\epsilon} _1$ from Eq.~(\ref{Eq:H0_3level}), which is shown in Fig.~\ref{fig:3-level_diff_diffk}. This example demonstrates that even when the Hamiltonian whose first non-zero term in the switching theorem is $\bar{\epsilon}_2$, the true error follows $1/T$ until $T$ is large enough so that the first non-trivial term dominates the asymptotic series.

\begin{figure}[t]
    \centering
    \includegraphics[width=0.49\textwidth]{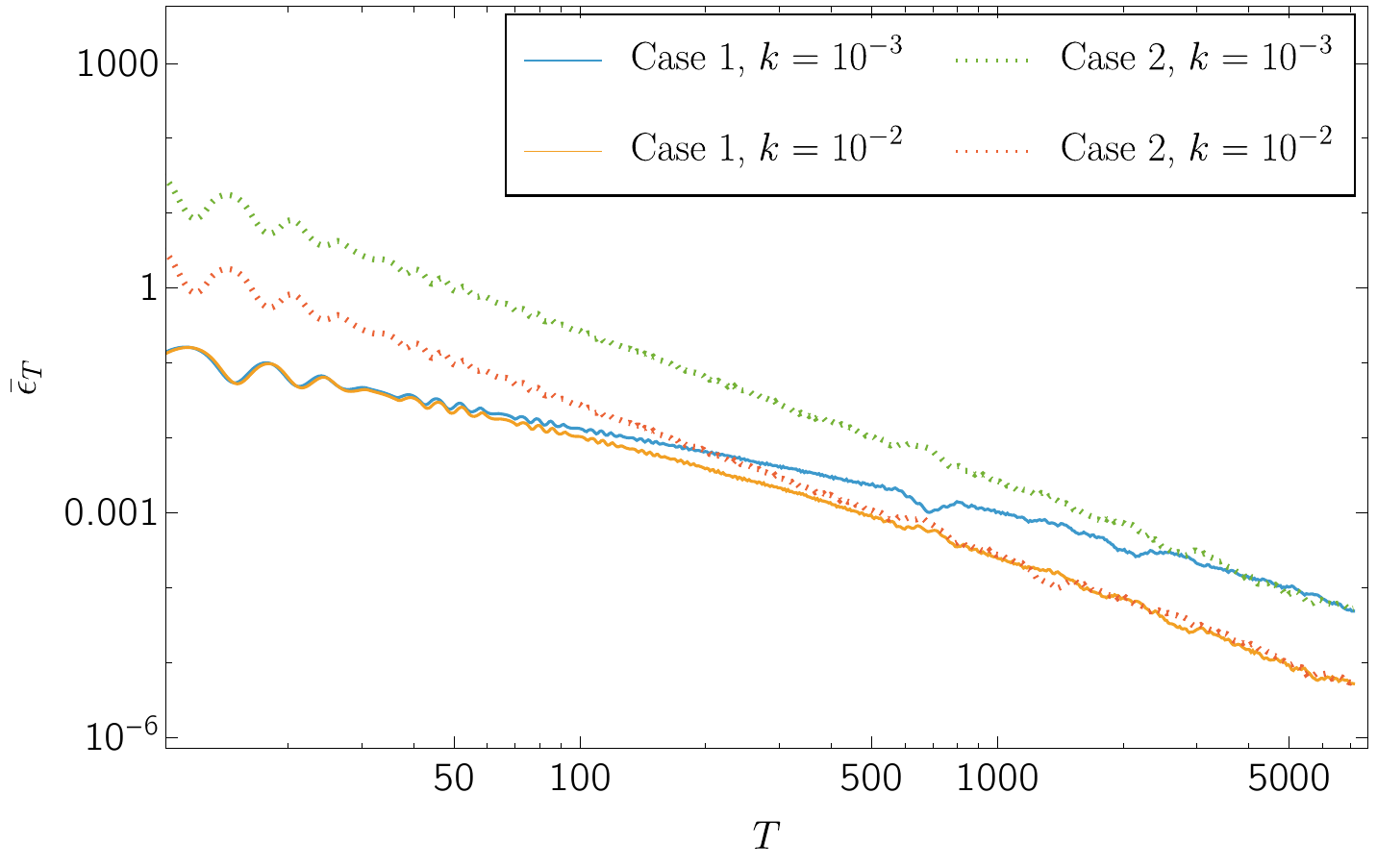}
    
    \caption{Comparison of Case 1 and Case 2 in Eqs.~(\ref{Eq:H_3levelex1}) and~(\ref{Eq:H_3levelex2}). Case 1 and 2 are represented by solid and dotted lines, respectively.
    }

    \label{fig:3-level_case12}
\end{figure}

We now compare the average true errors of Case 1 and Case 2 with $k=k_1$ and understand how modifications to the excited energies of the Hamiltonian affect the errors in the preparation of the final ground state. 
Fig.~\ref{fig:3-level_case12} shows that they are very similar in the hyperadiabatic regime.
It demonstrates that modifications of transitions between excited states are less likely to change the overall scaling of error in the hyperadiabatic regime. Furthermore, the transition between the ground state and the first excited state dominates the error in the hyperadiabatic regime, which is evident from Eq.~(\ref{Eq:barbn}). However, the figure shows that other transitions can affect the error for small $T$.

\section{Extreme Scenario: Essential Singularity at the endpoints }
\label{Sec:Exp}

\begin{figure}[t]
    \centering
    \includegraphics[width=0.49\textwidth]{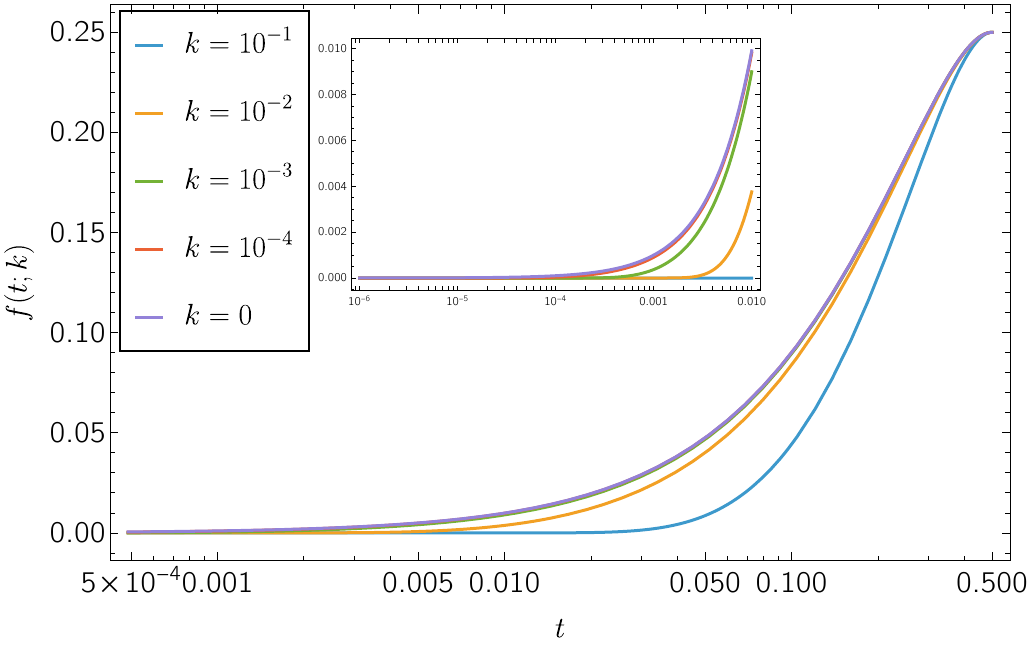}
    \caption{
    $f_{\exp}(t;k)$ with various $k$. $k=0$ corresponds to $f_0(t)$, whose first-order derivatives are not zero at the endpoints. The plot inside displays the behavior of $f_{\exp}(t;k)$ near the endpoint.
    }

    \label{fig:f_exp}
\end{figure}

\begin{figure}[t]
    \centering
    \includegraphics[width=0.49\textwidth]{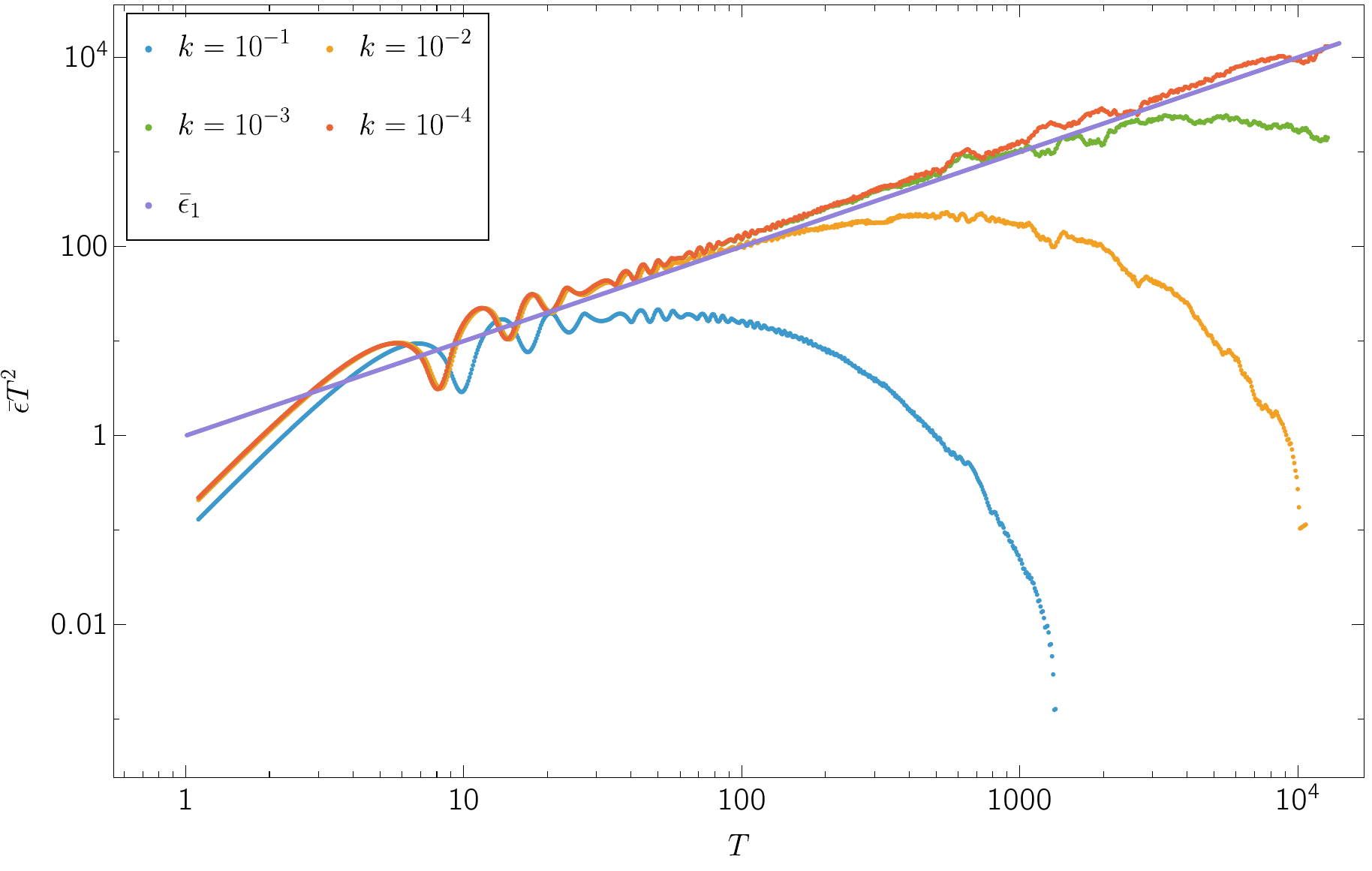}
    
    \caption{Average true errors are presented for various values of $k$ in Eq.~(\ref{Eq:exp1}) with $\bar{\epsilon}
    _1$ from the Hamiltonian at $k=0$. 
    }

    \label{fig:exp}
\end{figure}

In the previous examples, we tweaked the Hamiltonians such that the first derivatives of them were zero at the endpoints, while the intermediate evolution remained approximately the same. We saw that in the hyperadiabatic regime, the evolution switched to $T^{-2}$ scaling as expected. However, the agreement of the true error with the asymptotic scaling behavior widened for small $T$ as the differences in the intermediate evolution between the modified and original Hamiltonians became negligible. We could continue the process of refining the Hamiltonian evolution at its endpoints in a similar manner to make its higher-order derivatives zero vanish at the endpoints. The asymptotic error scaling suggests a faster decay with a more negative power of $T$ as the order of the lowest non-vanishing derivative at the endpoints increases. We could in particular consider the most extreme modification to the endpoints, which, mathematically speaking, amounts to introducing an essential singularity. This corresponds to a scenario where all the derivatives are zero at the end-points. An example of such a modification for the two-level Hamiltonian discussed in Section~\ref{Sec:Two} is shown below:
\begin{eqnarray}  
    H(t; k) & = 
    \begin{bmatrix}
    0 & E_1 f_{\text{exp}}(t; k)  \\
    E_1 f_{\text{exp}}(t; k)  & E_0
    \end{bmatrix}, \label{Eq:exp1}  \\ 
    \text{with } f_{\text{exp}}(t; k) & = 
        \begin{cases}
            t(1-t) \exp\left(-\frac{k}{t(1-t)}\right), & \text{if $0 \leq t \leq 1$} \\
            0, & \text{otherwise} \nonumber
        \end{cases}. 
\end{eqnarray}

As $k$ approaches zero, the difference between $H(t,k)$ in Eq.~(\ref{Eq:exp1}) and $H_{0}(t)$ in Eq.~(\ref{Eq:H0_2level}) at all intermediate times diminishes. The switching theorem is not particularly useful in estimating the error for the Hamiltonian in Eq.~(\ref{Eq:exp1}) when $k\neq 0$ since its derivatives are zero at the endpoints. One can always calculate the derivatives of $H$ to see that all the derivatives become zero at the endpoints. We cannot use the switching theorem here to make estimates for the errors. However, we can draw inspiration from the switching theorem to motivate such a modification to the Hamiltonian at its endpoints. This can be done by viewing the $f(t,k)$ as a limit of analytic functions with well-defined derivatives at the endpoints. The function $f_{\rm exp}(t,k)$ defined in Eq.~(\ref{Eq:exp1}) is compared against $f_{0}(t)$ defined in Eq.~(\ref{Eq:f0}) in Fig.~\ref{fig:f}. We compare the final error after time evolution with $H(t)$ from Eq.~(\ref{Eq:exp1}) with $k=10^{-4}\,, \, 10^{-3}\,, \, 10^{-2}$ to the error accumulated during the Hamiltonian evolution governed by Eq.~(\ref{Eq:H0_2level}).

Similar to previous examples, when $k$ is small, the Hamiltonian is nearly the same as Eq.~(\ref{Eq:H0_2level}), therefore its errors can be compared with $\bar{\epsilon}_1$ using the Hamiltonian at $k=0$, which is displayed in Fig.~\ref{fig:exp}\footnote{We note that the Schrodinger equation for Eq.~(\ref{Eq:exp1}) are solved with the initial point at $10^{-6}$ and the final point at $1-10^{-6}$, not 0 and 1, since the Hamiltonian is not well-defined numerically near the endpoints.}. $E_0$ and $E_1$ are chosen 1, as the example in Sec.~\ref{Sec:Two}. Fig.~\ref{fig:exp} shows that while all the derivatives of Eq.~(\ref{Eq:exp1}) is zero, when the timescale is small, the true errors for various $k$ follow the estimate from Eq.~(\ref{Eq:H0_2level}). However, the true errors end up following the asymptotic behavior, decaying faster than any power law as $T$ increases. Additionally, as in previous examples, decreasing $k$ delays the onset of the asymptotic region, despite the derivatives at the endpoints being zero in this case. However, for small $k$ and not so large values of $T$, $\bar{\epsilon}_T$ does not show features of exponential decay.

\section{Discussion} \label{Sec:Discussion}

The practical utility of the asymptotic scaling expression from the switching theorem depends on the timescale at which it becomes a reasonable approximation of the true error. In Section~\ref{Sec:Examples}, we compared switching theorem estimates with the true error in certain model Hamiltonian evolutions and observed that their agreement improves at larger timescales, or equivalently, in the hyperadiabatic regime.

The switching theorem states that for sufficiently long evolution times $T$, the errors propagated during adiabatic state preparation  exhibit characteristic power-law scaling in $1/T$ and can be estimated using only information about the Hamiltonian at its endpoints, independent of its intermediate evolution. This raises the question of whether modifying the Hamiltonian's rate of change at the initial and final times---while keeping the intermediate evolution largely unchanged---can reduce errors as predicted by the switching theorem. Mathematically, this corresponds to enforcing the vanishing of a few lower-order derivatives of the Hamiltonian at the endpoints. In this paper, we studied a few simple models with two-level and three-level Hamiltonians where such questions can be quantitatively asked. We found that the asymptotic scaling in $1/T$ is indeed the same as suggested by the switching theorem; however, the details of the behavior at the endpoints determine how quickly this asymptotic regime is reached.

As discussed in Section~\ref{Sec:Examples}, the applicability of the switching theorem for adiabatic state preparation is questionable. The best scenario for estimating errors is when the first non-trivial term provides a good approximation from relatively small timescales. However, if the timescale for the switching theorem to be useful is too large or the typical errors at such asymptotic limits are already too small, its utility is limited. In other words, the switching theorem is only practical when the timescale at which its first non-trivial term becomes a good approximation is small, and the errors near the onset of asymptotic behavior are sufficiently large to allow meaningful error estimation through scaling.

Although the numerical demonstrations in this work utilized low-dimensional models, the fundamental trade-off identified is expected to be universal for gapped quantum systems. The switching theorem derives its error estimates primarily from the behavior of the Hamiltonian at the endpoints, where ground and low-lying states contribute most significantly, rather than the specific dimensionality of the Hilbert space. Consequently, the `practical limitations' observed in our simple models are not artifacts of low dimensionality but are intrinsic features of the switching theorem, implying that more complex, realistic many-body Hamiltonians will confront the same barriers to efficient error reduction.

Let us consider a scenario where fault-tolerant analog quantum simulators are used, eliminating concerns about Trotterization errors. Even if all the other sources of errors are controlled, in order to estimate errors using the switching theorem, it requires information about the Hamiltonian at its endpoints. In realistic adiabatic evolution setups, the switching on and off of the Hamiltonian at the initial and final endpoints often involves sharp impulses, which can be challenging to manipulate efficiently. While these impulses may alter the true error from the switching estimate obtained using a desired evolution path, we find that tactful manipulations of the behavior at the endpoints can reduce the error by exploiting the switching theorem.    

On the other hand, in fault-tolerant digital quantum simulators, adiabatic state preparation must also account for Trotterization errors. This makes direct error estimation using the switching theorem impractical, necessitating further investigation into how Trotterization errors affect the switching theorem's predictions.


In this paper, we have considered the case where the auxiliary parameter $k$ is small, i.e., the new Hamiltonian path has zero first-order derivatives at the endpoints while the overall path is nearly the same as the original one. This is because we assume that the Hamiltonian path is given. 

While there are infinitely many Hamiltonian paths from the initial to the final Hamiltonians, not every path is applicable or efficient for adiabatic state preparation. For example, if there are phase transitions along the path~\cite{PhysRevLett.104.020502, PhysRevLett.104.207206, altshuler2009adiabaticquantumoptimizationfails, PhysRevA.80.062326}, adiabatic state preparation for this path will be hardly applicable. Therefore, Hamiltonians modified near the endpoints while not changing the overall path are examined.

Since it is expensive to prepare the final ground state with very small errors through adiabatic state preparation in practice, it may be more efficient to prepare an approximate final ground state using the adiabatic theorem, then use other methods to pin down the final state, for example, projection algorithms~\cite{Kitaev:1995qy, Choi:2020pdg, Stetcu:2022nhy, Cohen:2023rhd, Lin:2020zni, Dong:2022mmq, Poulin:2008mek, Ge:2019shk}. In this case, the scaling behavior predicted by the switching theorem is of limited practical use, as it applies only in the regime of extremely large timescales, where diabatic errors are already negligible.

\section*{Author contributions}

All authors contributed significantly and in complementary ways to this work. AL conducted the initial study of the systems as part of a high school summer project under the guidance of HO, MP, and TC. HO drafted the initial version of the manuscript, generated the figures, and participated in manuscript revisions. MP supervised the project, contributed to writing, and assisted in editing. TC oversaw the project throughout all stages. The contributions were balanced, with each author playing a key role in the completion of the study.

\section*{Data availability statement}

Data will be made available on reasonable request.

\begin{acknowledgments}

This work was supported in part by the U.S. Department of Energy, Office of Nuclear Physics under Award Number(s) DE-SC0021143, and DE-FG02-93ER40762.

\end{acknowledgments}
  
\bibliography{refs.bib}

\end{document}